\documentclass[twocolumn,runningheads]{svjour2}
\smartqed  
\usepackage{graphicx}
\usepackage{mathptmx}      
\journalname{Astrophysics and Space Science}
\begin{document}

\title{Synchrotron emission from secondary leptons in microquasar jets
}



\author{P. Bordas         \and
        J.~M. Paredes  \and
        V. Bosch-Ramon \and
        M. Orellana  
}

\authorrunning{P. Bordas, J.~M. Paredes, V. Bosch-Ramon, M. Orellana
} 

\institute{P. Bordas, and J.~M. Paredes \at
              Departament d'Astronomia i Meteorologia,
              Universitat de Barcelona,
              Mart\'{\i} i Franqu\`es 1,
              08028, Barcelona, Spain\\
              Tel.: +34-934039225, +34-934021130\\
              Fax: +34-934021133\\
              \email{pbordas@am.ub.es; jmparedes@ub.edu}  
           \and
           V. Bosch-Ramon \at
              Max Planck Institut f\"ur Kernphysik,
              Saupfercheckweg 1,
              Heidelberg 69117, Germany\\
              Tel.: +49 6221516586\\
              Fax: +49 6221516324\\
              \email{vbosch@mpi-hd.mpg.de}  
           \and
           M. Orellana \at
              Instituto Argentino de Radioastronom\'ia, CC5, (1894) Villa Elisa, Buenos Aires, Argentina. \\
	      Tel.: +54-221-482-4903\\
	      fax:  +54-221-425-4909\\
	      \email {morellana@irma.iar.unlp.edu.ar} \\
}

\date{Received: date / Accepted: date}

\maketitle

\begin{abstract}

We present a model to estimate the synchrotron radio emission generated in microquasar (MQ) jets 
due to secondary pairs created via decay of charged pions produced in proton-proton collisions 
between stellar wind ions and jet relativistic protons. Signatures of electrons/positrons are obtained 
from consistent particle energy distributions that take into account energy losses due to synchrotron and inverse 
Compton (IC) processes, as well as adiabatic expansion. The space parameter for the model is explored 
and the corresponding spectral energy distributions (SEDs) are presented. We conclude that secondary 
leptonic emission represents a significant though hardly dominant contribution to the total radio 
emission in MQs, with observational consequences that can be used to test some still unknown
processes occurring in these objects as well as the nature of the matter outflowing in their jets.

\keywords{microquasars \and radio emission  \and secondary leptons}
\PACS{98.38.Fs \and 97.80.Jp \and 95.30.Cq}
\end{abstract}

\section{Introduction}
\label{intro}

X-ray binary systems (XRBs) are composed by either a stellar mass black hole or a neutron star, 
and a normal (non degenerated) star which supplies matter to the compact object through the formation
of an accretion disk. Some ~260 XRBs are known up to now \cite{liu06} probably corresponding to an underlying 
population of some tens of thousands of compact objects in our Galaxy. A few of these sources 
present also non-thermal radio emission, hence evidencing the existence of mechanism(s) capable
of injecting and/or accelerating large numbers of relativistic particles. Some radio emitting 
X-ray binary systems (REXBs) have been observed showing ejection of material at relativistic 
velocities and to display jets like those seen in quasars and active galactic nuclei but 
at $\sim 10^{-6}$ times shorter scales. This analogy is the reason for calling them microquasars (MQs) 
\cite{mirabel99} and making them some of the most interesting objects for astrophysics.
Furthermore, attention on these objects has grown since the proposal of Paredes et al. (2000)
\cite{paredes00} of MQs as counterparts of some of the unidentified gamma-ray sources of the EGRET
catalog \cite{hartman99} and hence pointing them as plausible high energy emitters. A strong 
confirmation of this association has come from the detections of the MQs LS 5039 and LS I +61 303 at
Tev energies using respectively the ground-based Cherenkov telescopes HESS \cite{aharonian05} and 
MAGIC \cite{albert06}, giving support and empowering at the same time a number of previous detailed
studies centered on the mechanisms operating in these sources in order to explain the gamma ray domain 
(see, e.g., \cite{bosch05} and \cite{romero05}). Moreover, a jet origin of the emission 
from MQs has been suggested from the observation of syncrothron emission of relativistic 
electrons/positrons extending from the radio all the way into the X-ray regime. 
In this sense jet-like models have focused on different approaches regarding the particle 
origin that could generate the required emission properties in a consistent way. Some of 
them consider leptons directly injected at the base of the jet and, in extending outwards, 
Compton-interaction with external/self-created photon fields produces high energy radiation. 
Other models deal with an hadronic origin of the high energy emission, through proton-proton 
interactions and pion decay producing  gamma rays and leaving the resulting co-generated leptons
as low energy emitters. The present work refers to the later procedure, focusing on the modelisation
of the secondary leptonic synchrotron emission in order to constrain the characterization of MQ 
jets. An outline of the model is given in the next, followed by the results showing the SEDs and lightcurves
under different parameter assumptions and the conclusions that can be extracted from them.

\section{Model description}

We have taken a binary system formed by a black hole and a high-mass 
early-type star which feeds the accretion transfer of mass onto the compact object while developing 
an accretion disk. Part of the accretion power is 
converted to kinetic and magnetic energy of the accretion flow under the effects of 
the compact object potential well and a relativistic $e^{\pm}$-$p$ plasma is ejected in a 
direction taken to be perpendicular to the plane defined by the accretion disk. This
plane is the same as the orbital one, even if this condition can be relaxed to allow
a more general situation but for simplicity we assume coplanarity in our model. 
Jet energetics is assumed to be dominated by accretion, and further energy sources like 
compact object rotation have been neglected at this stage. The jet will also contain a 
magnetic field $B_{jet}$ associated with the plasma. We assume that the matter kinetic luminosity 
is higher than the magnetic luminosity (or total magnetic energy crossing the jet section per 
time unit) in the jet regions we are concerned with, although the magnetic field can be still 
significant once the jet is formed, since the ejection mechanism have likely a 
magneto-hydrodynamical origin.

In this scenario we deal with radiative processes that take place in the jet and can produce 
significant emission in the radio spectrum. Although the contribution of protons could still 
be significant from the radiative point of view, we will focus on the leptonic component only. 
The reader is related to other works (see, e.g., \cite{romero03} and \cite{romero05}) for treatments 
on primary hadrons.

\begin{table}[t]
\caption{Parameter values used throughout the model.}
\centering
\label{parameters}
\begin{tabular}{lcc}
\hline\noalign{\smallskip}
parameter & Symbol & Value\\
\tableheadseprule\noalign{\smallskip}
  & $$ & $$\\
Black hole mass  & $M_{bh}$ & 3$M_{\odot}$ \\
Injection point  & $z_{0}$ & $50R_{g}$\\
Initial radius  & $R_{0}$ & $5R_{g}$\\
Radius of the companion star  & $R_{\star}$ & $15R_{\odot}$\\
Orbital radius & $a$ & $3R_{\star}$\\
Luninosity companion star  & $L_{\star}$ & $1.6\times10^{39}$ erg/s\\
Mass loss rate  & $\dot{M}_{\star}$ & $3\cdot10^{-6}M_{\odot}$yr$^{-1}$\\
Jet's Lorentz factor & $\Gamma$ & $1.02$\\
Jet kinetic luminosity  & $Q_{j}$ & $10^{36}$ erg/s\\
Proton kinetic Luminosity  & $Q_{p}$ & $10^{35}$ erg/s\\
Minimum proton energy  & $E_{p}^{min}$ & $2$ GeV\\
Maximum proton energy  & $E_{p}^{max}$ & $100$ TeV\\
Leptonic spectral index & $P$ & $2.2$\\
Wind-jet penetration factor  & $f_{p}$ & $0.1$\\
Magnetic field at $z_{0}$  & $B_{0}$ & $10^{2}$,$10^{3}$, $10^{4}$ \\
  & $$ & $$\\

\noalign{\smallskip}\hline
\end{tabular}
\end{table}

\subsection{{\it Secondary generation}}

Secondary leptons are generated in $p$-$p$ interactions of hot protons in the jet with the 
companion star wind that extents out isotropically producing charged and neutral pions
($\pi^{\pm}$, $\pi^{0}$)through the reaction channel 
$p+p\rightarrow p+p+\xi_{\pi^{0}}\pi^{0}+\xi_{\pi^{\pm}}(\pi^{+}+\pi^{-})$, 
where $\xi_{\pi^{\pm}}\sim2(E_{p}/GeV)^{1/4}$
is the charged pion multiplicity. The relativistic injection proton spectrum is a
power law $N_p(E_p)= K_p E_p^{-\alpha}$ where $E_{p}^{min}\leq E_p\leq E_{p}^{max}$ where 
the constant $K_p$ can be found normalisation to the total power that goes to protons $Q_p$.
The corresponding proton flux will be given by
$J_p(E_p)=(c/4\pi)K_p(z_0/z)^{2}E_p^{-\alpha}$. We assume that a fraction of $1/10$ of the 
matter that crosses the jet region penetrates into
it. When charged and neutral pions are created, the first will decay to muons 
and subsequently to electrons and positrons, while the second decays to high energy photons through

$\space$

$\pi^{+}\rightarrow\nu_{\mu}+\mu^{+}\rightarrow\nu_{\mu}+e^{+}+\nu_{e}+\bar{\nu}_{\mu}$

$\pi^{-}\rightarrow\bar{\nu}_{\mu}+\mu^{-}\rightarrow\bar{\nu}_{\mu}+e^{-}+\bar{\nu}_{e}+\nu_{\mu}$ 

$\pi^{0}\rightarrow2\gamma $.

$\space$

\noindent For an injection proton spectrum as the one given above, the pion spectrum (in the jet reference
frame) will be a power law and the electron/positron distribution will also follow a power
law \cite{ginzburg64} with a differential pair injection rate given by
$J_e (E_e )= K_e E_e^{-P}$ where we use a value $P=2.2$.

\subsection{{\it Secondary evolution}}

This injected electron-positron population suffers subsequent radiative cooling due to synchrotron
and inverse Compton losses as well as adiabatic expansion given by

$\space$
 
$\left[\frac{dE_{e^{\pm}}}{dt}\right]_{sync}=-2.36\cdot10^{-3}B^{2}E_{e^{\pm}}^{2}$, 

$\left[\frac{dE_{e^{\pm}}}{dt}\right]_{IC}=-3.90\cdot10^{-2}U_{star}E_{e^{\pm}}^{2}$,

$\left[\frac{dE_{e^{\pm}}}{dt}\right]_{exp}=\frac{2}{3}\frac{V_{exp}}{R(z)}E_{e^{\pm}}$ 

$\space$

\noindent respectively. 
Here $U_{star}$ is the companion star's photon field given by $U_{star}=\frac{L_{\star}}{4\pi d^{2}c}$ where 
$d$ is the distance from the star and in our case $L_{\star}=1.6\cdot10^{39}$ erg/s, and $V_{exp}= dR(z)/dt$ 
is the lateral 
expansion velocity of the jet. Fresh electrons and positrons that have been injected at a 
certain point into the jet suffer the different energy losses modifying in this way the spectral 
distribution of particles at each height $z$. Moreover, different evolution stages sum at each 
height since injection of fresh particles occurs all along the jet and therefore a mix of multiple 
evolved population of electron/positron distribution arises. To compute it in a consistent way, 
one has to assume the continuity equation, 

$\space$

$N_{e^{\pm}}(E_{e^{\pm}},\, z)dE_{e^{\pm}}=N_{0,e^{\pm}}(E_{0,e^{\pm}},\, z_{0})dE_{0,e^{\pm}}$

$\space$

\noindent where $N_{0,e^{\pm}}(E_{0,e^{\pm}},\, z_{0})=K_{0,e^{\pm}}E_{0,e^{\pm}}^{-p}$ is the initial energy
spectrum for the injected particle density. Solving the equations for the evolution of the particle 
energy along the jet axis, one can find
the spectral distribution at each slice, compute de differential luminosity at each height and 
finally integrate to find the total luminosity.

\begin{figure}
\centering
\includegraphics[width=0.45\textwidth]{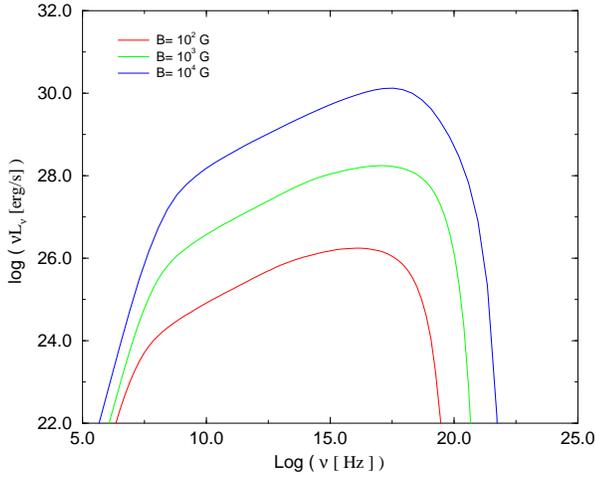}
\caption{The strength of the magnetic field $B$ is treated as a free parameter in our model. Synchrotron loses will
be enhanced when taking higher values for $B$ but, since injection occurs all along the jet inside the
bynary system (until a distance roughly given by $z\simeq R_{orb}\sim 10^{13}$ cm), the enhancement in the final emission
will preval over losses. Here we show the SEDs for three different magnetic fields (B); 
values indicated are taken at the base of the jet.}
\label{figcurve} 
\end{figure}

\begin{figure}
\centering
\includegraphics[width=0.45\textwidth]{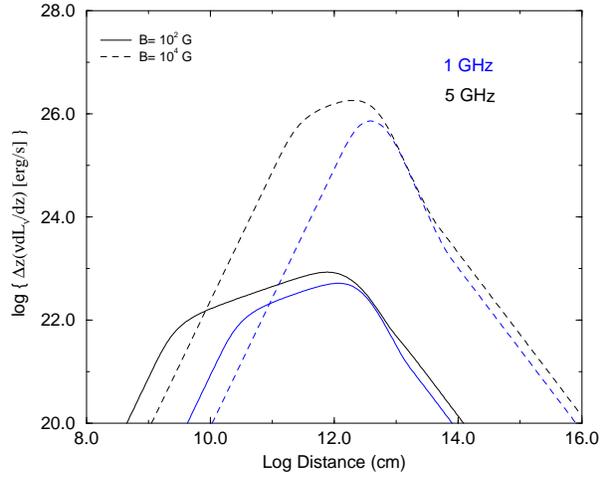}
\caption{Light curves of emission at 1 and 5 GHz bands at growing distances along the jet for 
various magnetic fields $B$ wich values are refered to the base of the jet at $z_{o}= 50 R_{g}$. Not only the
overall emission but also the peak of the lightcurves moves to higher distances when increasing $B$, as well as
the difference between the locations of the peaks for the two frequencies.}

\label{figsed} 
\end{figure}

\begin{figure}
\centering
\includegraphics[width=0.45\textwidth]{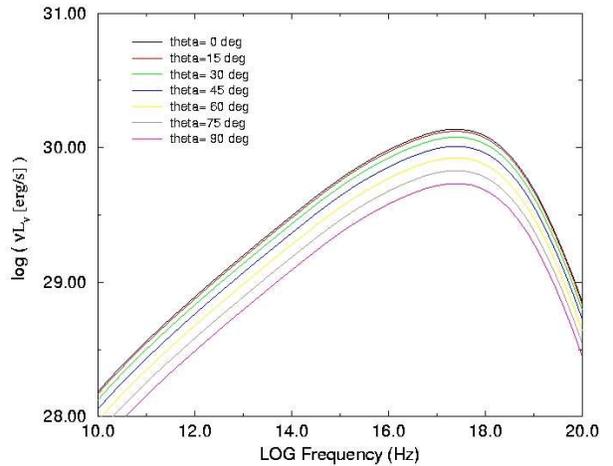}
\caption{Variations of the SEDs due to different observing angles $\theta$  between the approaching 
jet axis and the observer line of sight.As far as the jet material travels at higher velocities 
along the jet, the boosting will get increasingly greater while observing at smaller angles. 
The SEDs showed here correspond to $\theta$  ranging from 0 to $\frac{\pi}{2}$ , where we have 
used a Lorenz factor $\Gamma\sim1.02$ corresponding to a mildly relativistic jet with 
$v_{j}\sim0.2c$ } 
\end{figure}

\subsection{{\it Secondary emission}}
 
Secondary electrons and positrons will radiate through synchrotron process since we assume 
the presence of a magnetic field wich is tangled to the plasma and oriented randomly in 
direction. The resulting emission will be isotropic in the jet reference frame; 
to compute the expected luminosity, we have used the formulae presented in \cite{pacholczyk70}, 
where expressions for the specific emission and absorption coefficients
are given by

$\space$

$\varepsilon_{\nu}=C_{3}B\int_{0}^{\infty}N_{e^{\pm}}(E_{e^{\pm}})F(x)dE_{e^{\pm}}$

$k_{\nu}=-\frac{c^{2}}{2\nu^{2}}C_{3}B\int_{0}^{\infty}E_{e^{\pm}}^{2}\frac{d}{dE}\left(\frac{N_{e^{\pm}}
(E_{e^{\pm}})}{E_{e^{\pm}}^2}\right)F(x)dE_{e^{\pm}}$
 
$\space$

\noindent with the function F(x) given by $F(x)=x\int_{x}^{\infty}K_{5/3}(z)dz$ being $K_{5/3}$ the modified
second kind Bessel functions, $x\equiv\frac{\nu}{\nu_{c}}$ where
$\nu_{c}=\frac{3e}{4\pi m_{e}^{3}c^{5}}E_{e^{\pm}}^{2}\approx6.27\cdot10^{18}E_{e^{\pm}}^{2}$ Hz is
the critical frequency, and the constant $C_{3}\approx1.87\cdot10^{-23}$ (in CGS units). 
Then, the specific differential luminosity at a certain heigh z in the jet can be expressed as

$\space$

$\frac{dL_{\nu}(z)}{dz}=2\pi R(z)\frac{\varepsilon_{\nu}}{k_{\nu}}\left(1-e^{-l_{j}k_{\nu}}\right)$

$\space$

\noindent where $l_{j}$ is the typical size of the synchrotron emitting plasma region. 
Now, integrating over the entire jet lenght we find the spectral energy distribution, 

$\space$

$\nu L_{sync}=\nu\int_{z_{0}}^{z_{max}}\delta^{2}\frac{dL_{\nu}(z)}{dz}dz$

$\space$

\noindent where $\delta=\left[\Gamma\left(1-\beta\cos\theta_{obs}\right)\right]^{-1}$
is the Doppler boosting factor.

\section{Model results and conclusions}

SEDs are obtained for different magnetic field values, 
electron/positron spectral indices and spatially distributed disks. We have estimated 
also the expected emission along the jet at 1 and 5 GHz. Leptons are injected in the 
context of hadronic secondaries generation within a detailed model that takes into 
account in a consistent way particle injection mechanisms and cooling due to radiation 
processes and adiabatic expansion. 
The luminosities obtained are slightly lower than in the models based on primary leptons 
injection, and must be considered complementary to them. However, we note that within our 
model there is no requirements of acceleration processes along the jet to obtain the final emission 
results. Such acceleration processes are still not well understood, although. They 
could come from diffusive shock acceleration along the jet when fresh ejecta interact with previous 
blobs of plasma already outflowing at lower velocities. Other scenarios assume a continuous 
energy transfer mechanism from the magnetic field to the matter content of the jet in such 
a way that the resulting parsec-scale radio emission can be explained. The fact of studying 
alternative models were particles are directly injected until a certain height along the jet 
can constrain the amount of acceleration required and contribute to the understanding of the 
physical mechanisms that can lead to such processes.

Signatures at different distances along the jet and specific spectral features detectable 
for reasonable parameter values treated in our numerical simulations have the potential to 
be an important clue for determining the matter content of jets. In particluar, highly 
resolved observations at 1 and 5 GHz could determine if leptons are present at heights $10^{12-13}$ 
cm at the edge of the binary system typical region where wind matter from the companion 
is still significant. If electrons/positrons still show high energies due to a recent injection 
from hadronic interactions at these parts of the jet, it could be a signature of secondary 
generation without the necessity of invoking additional acceleration processes.

\begin{acknowledgements}

P.B and J.M.P acknowledge support by DGI of the Spanish Ministerio
de Educaci\'on y Ciencia (MEC) under grant
AYA2004-07171-C02-01, as well as partial support by the European Regional
Development Fund (ERDF/FEDER).
V.B-R. thanks the Max-Planck-Institut f\"ur Kernphysik for its support and
kind hospitality.
M.O is supported by CONICET (PIP 5375) and ANPCyT (PICT 03-13291), Argentina.
\end{acknowledgements}

\end{document}